\documentclass{aa}
\usepackage{graphicx}
\usepackage{lscape}

\begin{document}
\def\lsim{\, \lower2truept\hbox{${<\atop\hbox{\raise4truept\hbox{$\sim$}}}$}\,}
\def\gsim{\, \lower2truept\hbox{${>\atop\hbox{\raise4truept\hbox{$\sim$}}}$}\,}
   \title{High-frequency radio observations of the K\"uhr sample
   and the epoch-dependent luminosity function of flat-spectrum quasars}
\titlerunning{High-frequency radio observations of the K\"uhr
sample}
%   \subtitle{I. Overviewing the $\kappa$-mechanism}

\author{R. Ricci\inst{1,2}
          \and
          I. Prandoni\inst{3}
          \and
          C. Gruppioni\inst{4}
          \and R.J. Sault\inst{2}
          \and G. De Zotti\inst{5,1}
          }

\offprints{R. Ricci}

\institute{SISSA/ISAS, Via Beirut 2--4, I-34014 Trieste, Italy \\
         \and
Australia Telescope National Facility, CSIRO, P.O. Box 76, Epping,
    NSW 2121, Australia \\ \email{Roberto.Ricci@atnf.csiro.au}
\and INAF, Istituto di Radioastronomia, Via Gobetti 101,
I--40129, Bologna, Italy \\
\and INAF, Osservatorio Astronomico di Bologna, Via Ranzani 1,
I--40126 Bologna, Italy \\
\and INAF, Osservatorio Astronomico di Padova, Vicolo
dell'Osservatorio 5, I--35122 Padova, Italy \\
\email{dezotti@pd.astro.it}
             }

\date{Received ...... ; accepted .......}

\abstract{We discuss our ATCA 18.5 and 22 GHz flux density
measurements of Southern extragalactic sources in the complete 5
GHz sample of K\"uhr et al. (1981). The high frequency (5--18.5
GHz) spectral indices of steep-spectrum sources for which we have
18.5 GHz data (66\% of the complete sample) are systematically
steeper than the low frequency (2.7--5 GHz) ones, with median
$\alpha^5_{2.7} = 0.76$,  median $\alpha^{18.5}_{5} = 1.18$
(S$_{\nu}\propto \nu^{-\alpha}$), and
median steepening $\Delta\alpha = 0.32$, and there is evidence of
an anti-correlation of $\Delta\alpha^{18.5}_{5}$ with luminosity.
The completeness of 18.5 GHz data is much higher (89\%) for
flat-spectrum sources (mostly quasars), which also exhibit a
spectral steepening: median $\alpha^5_{2.7}=-0.14$, median
$\alpha^{18.5}_{5}=0.16$ (S$_{\nu}\propto \nu^{-\alpha}$), 
and median $\Delta\alpha = 0.19$. 
Taking advantage of the almost complete redshift information on
flat-spectrum quasars, we have estimated their 5 GHz luminosity
function in several redshift bins. The results confirm that their
radio luminosity density peaks at $z_{\rm peak} \simeq 2.5$ but do
not provide evidence for deviations from pure luminosity evolution
as hinted at by other data sets. A comparison of our 22 GHz flux
densities with WMAP K-band data for flat-spectrum sources suggests
that WMAP flux densities may be low by a median factor of $\simeq
1.2$. The extrapolations of 5 GHz counts and luminosity functions
of flat-spectrum radio quasars using the observed distribution of
the 5--18.5 spectral indices match those derived directly from
WMAP data, indicating that the high frequency WMAP survey does not
detect any large population of FSRQs with anomalous spectra.

   }

   \maketitle
%
%________________________________________________________________
\section{Introduction}

High radio frequency (10-100 GHz) sky surveys have started to become 
feasible only very recently. The first all-sky surveys at high
radio frequencies have been provided by the WMAP satellite (Bennett et al. 
2003). In particular, the complete sample of extragalactic sources drawn 
from its $K$-band catalog ($S \ge 1.25\,$Jy, 
$|b|\ge 10^\circ$; see De Zotti et al. 2005) comprises 155 extragalactic 
sources (plus a planetary nebula), over an area of 10.4 sr. 

Another survey which which will give a significant contribution to the 
knowledge of the $>10$ GHz extragalactic sky is the ongoing 
all-sky 20 GHz survey which is being undertaken with the Australia Telescope
Compact Array (ATCA) in the Southern emisphere (a pilot survey covers about
1200 deg$^2$ to 100 mJy, Ricci et al. 2004b).

An example of a deeper survey is the 9th Cambridge survey 
carried out at 15 GHz with the Ryle Telescope (Waldram et al. 2003), which 
covers 520 deg$^2$ to $S=25$ mJy, and reaches deeper flux
densities on smaller areas. 

Such surveys are of fundamental importance to directly provide information 
about the extragalactic populations dominating the 
sky at $>10$ GHz frequencies, and allow to test extrapolated models based 
mainly on lower frequency selected samples. 

In this paper we present 18.5 and 22~GHz observations of the Southern 
extragalactic sources of the complete 5~GHz sample of K\"uhr et 
al. (1981), with the aim of better modeling the high frequency properties of 
radio galaxies and quasars, and check whether radio sources detected by WMAP
in the K-band significantly differ from the ones dominating at lower 
frequencies. 

\bigskip

In Sect.~2 we briefly present the 18.5 and 22 GHz flux density measurements.  
The main high frequency properties of Southern K\"uhr sources are analysed 
in Sect.~3. In Sect.~4 we estimate
the 5 GHz luminosity function (LF) of FSRQs at different cosmic
epochs and compare it with predictions of a recent model by De
Zotti et al. (2005). The observed 5--18.5 GHz spectral index
distribution is then exploited to extrapolate the LF to 22.8 GHz,
the central frequency of the WMAP K-band. The extrapolated LF is
compared with a direct estimate from the WMAP data. 
In Sect.~5 we summarize our main conclusions.

Throughout this paper we adopt a flat $\Lambda$ cosmology with
$\Omega_{\Lambda}=0.7$ and $H_0=70\,\hbox{km}\,\hbox{s}^{-1}\,
\hbox{Mpc}^{-1}$.

\section{The data} \label{sec:select}

In March 2002 we have carried out total intensity and linear
polarization measurements at 18.5 GHz of 249 of the 258 southern
($\delta <0^\circ$) extragalactic sources in the 5 GHz all-sky 1
Jy sample (K\"uhr et al. 1981; Stickel et al. 1994) with the
Australia Telescope Compact Array (ATCA), using the prototype
high-frequency receivers mounted on antennas CA02, CA03, and CA04.
The array configuration was quite compact, giving a $\hbox{HPBW}=
15''.6$. Full details on observations and data reduction are
provided by Ricci et al. (2004a, hereafter referred to as Paper I), 
who carried out an analysis
of linear polarization data. Many of these sources had already
been targeted with the ATCA as part of a search for potential 22~GHz 
calibrators. The remaining 22~GHz total intensity measurements 
were taken with ATCA in January 2001 using the three antennas of the 
750C array configuration providing a $\hbox{HPBW}=7''.0$. 

The 18.5 and 22 GHz flux densities are listed in Table 1, where, for 
each observed source, we also report the 5~GHz
flux density, the spectral index between 2.7 and 5 GHz (both from 
K\"uhr et al. 1981), the source type (from 
Stickel et al. 1994) and the source redshift (mostly from 
Stickel et al. 1994, with some additional redshifts from the NED
database). 

Both 18.5 and 22 GHz data were reduced as discussed in Paper~I. 
In particular, flux densities 
are derived using non-imaging model-fitting techniques, which assume a point 
source model. We notice that the 22 GHz data have larger uncertainties,
are less homogeneous and, having better angular resolution, can suffer
more from resolution effects. Therefore, our analysis mostly relies on
18.5 GHz data.

Whenever a source at 18.5 GHz is poorly modeled as a point source 
(see Paper~I
for details), such source is labeled ``R'' (for ``resolved'') in the last 
column of Table 1, and excluded from further analysis. 

Small differences with the 18.5 GHz flux densities, and
associated errors, reported by Ricci et al. (Paper~I), are due to a
new analysis of the measurements. The 18.5~GHz errors now include the flux
calibration uncertainty of 5\%. Calibration errors, estimated at 
10\%, are also included in the 22-GHz data.  

Following Stickel et al. (1994), we have classified as flat-spectrum, 
sources with 2.7--5 GHz spectral index $\alpha^5_{2.7} < 0.5$ ($S_\nu \propto
\nu^{-\alpha}$); sources with larger values of $\alpha^5_{2.7}$
are classified as steep-spectrum. Of the 249 sources observed, 146 are
flat-spectrum and 103 are steep-spectrum.

In the following analysis we exclude ``resolved'' (``R'') sources. This 
means 130 out of the 146
flat-spectrum sources in the original complete sample (89\%) and
68 of the 103 steep-spectrum sources (66\%) are retained.
Special emphasis will be given to Flat-Spectrum Radio Quasars (FSRQs), 
which constitute the population which is best represented in our sample.

\begin{figure}
   \centering
\includegraphics[height=7cm, width=8.5cm]{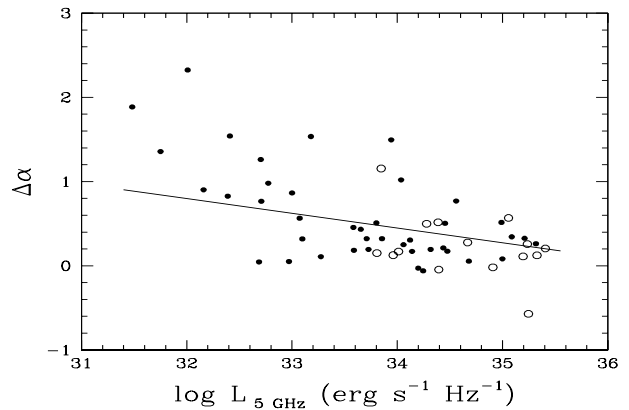}
\vskip-1.2cm \caption{High-frequency steepening of steep-spectrum
sources as a function of the 5 GHz luminosity. Radio galaxies and quasars 
are indicated by filled and empty circles respectively.
The regression line is $\Delta \alpha= -0.1749 \log L_{5{\rm GHz}} + 6.3948$. }
\label{fig:steep}
\end{figure}

\begin{figure}
   \centering
\includegraphics[height=10cm, width=8.5cm]{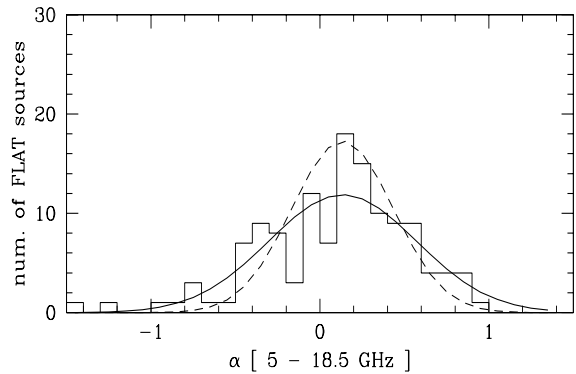}
\vskip-4.3cm\caption{Distribution of 5--18.5 GHz spectral indices,
$\alpha_5^{18.5}$, of flat-spectrum sources in the K\"uhr sample.
The solid line shows the formal best-fit Gaussian representation
of the distribution ($<\alpha>=0.134$, $\sigma=0.437$), the dashed
line is a Gaussian with the same $<\alpha>$ and $\sigma=0.30$ (see text). }
\label{fig:alphisto}
\end{figure}

\section{Data analysis} \label{sec:analysis}

Since the high frequency information on steep-spectrum sources is 
seriously incomplete (see above), and, as a consequence, any conclusion 
on their spectral properties may be biased, we only note that they show 
evidence of a spectral steepening. The median $\alpha^5_{2.7}$ is 0.76 
(S$_{\nu}\propto \nu^{-\alpha}$) 
(the mean is $0.81 \pm 0.02$), while the median $\alpha^{18.5}_{5}$ is 
1.18 (mean $1.28\pm 0.06$), and the median steepening is $\Delta\alpha =
0.32$ (mean $0.47\pm 0.06$). The steepening is anti-correlated
with luminosity (see Fig.~\ref{fig:steep}) or redshift (the
Spearman's rank correlation coefficient yields a probability of no
correlation of $3\times 10^{-4}$). This translates into an
anti-correlation of $\alpha^{18.5}_{5}$ with luminosity (or $z$),
possibly reversing the {\it positive} correlation with luminosity
of the low frequency spectral index (see, e.g., Dunlop \& Peacock
1990). Since low-luminosity (and low $z$) sources are mostly
galaxies, while quasars are high-luminosity (and high $z$) sources 
(see Fig.~\ref{fig:steep}), we have larger steepenings for galaxies (median
$\Delta\alpha = 0.34$) than for quasars (median $\Delta\alpha =
0.20$). 

A high-frequency steepening is also observed for flat-spectrum
sources, which are mostly FSRQs. The median values are
$\alpha^5_{2.7}=-0.14$ (S$_{\nu}\propto \nu^{-\alpha}$), 
$\alpha^{18.5}_{5}=0.16$, $\Delta\alpha =
0.19$, while the mean values are, respectively, $-0.13\pm 0.03$,
$0.11 \pm 0.04$, and $0.24\pm 0.05$. The best-fit Gaussian
distribution of 2.7--5 GHz spectral indices has a mean of
$-0.10$ and a dispersion of 0.30, while that of 5--18.5 GHz
spectral indices has mean 0.13 and dispersion 0.44
(Fig.~\ref{fig:alphisto}). The large value for the dispersion of
the distribution of $\alpha^{18.5}_{5}$ may be partly due to
measurement errors and to uncertainties in the source
classification; we will come back to this in
Sect.~\ref{sec:WMAPlf}. The Spearman's rank correlation statistics
detects a {\it positive} correlation of $\alpha^{18.5}_{5}$ with
luminosity (or $z$), although with a lower significance than for
steep-spectrum sources (probability of no correlation $2\times
10^{-3}$).

\subsection{Comparison with WMAP K-band flux measurements}
 
It is interesting to compare our 18.5~GHz flux density measurements of
flat-spectrum unresolved sources of the K\"uhr sample with those of WMAP 
in the K-band (22.8 GHz).
A cross-correlation with the WMAP catalog (Bennett et al. 2003)
yields 59 matches within angular separations of 
$11$ arcmin (the Bennett et al. 2003 criterion for WMAP source 
cross-identifications). For the
un-matched flat-spectrum sources in our sample we adopt an upper flux
density limit in the K-band of $S_K=1.25$ Jy, i.e. the completeness
limit of the WMAP survey, as derived by De Zotti et al. (2005; see
also Arg\"ueso et al. 2003). We notice, however, that 
Bennett et al. (2003) give a less
conservative estimate of $S_K=0.75$ Jy.  Using the Kaplan-Meier
estimator [routine KMESTM in the software package ASURV Rev 1.2
(Isobe \& Feigelson, 1990)] we get an
estimated median value of $\log(S_{18.5{\rm GHz}}/S_K) = 0.098$
(corresponding to $S_{18.5{\rm GHz}}/S_K = 1.25$). 
This includes WMAP flux density upper limits 
by implementing the survival analysis methods presented in Feigelson
\& Nelson (1985) and Isobe, Feigelson \& Nelson (1986). The mean value
of $\log(S_{18.5{\rm GHz}}/S_K)$, $0.126+/-0.021$, is less
reliable because it is significantly affected by outliers (probably 
variable sources). 

The median $\alpha^{18.5}_{5}=0.16$ would imply a significantly
lower value ($\log(S_K)=\log(S_{22.8{\rm GHz}})=\log(S_{18.5{\rm GHz}}) 
-0.16 \log(22.8/18.5)$, or $\log(S_{18.5{\rm GHz}}/S_K) = 0.015$) and using the
actually observed distribution of $\alpha^{18.5}_{5}$ for flat-spectrum 
sources (see Fig.~\ref{fig:alphisto}) we obtain an even lower mean value 
($\log(S_{18.5{\rm GHz}}/S_{22.8{\rm GHz}}) = 0.0097\pm 0.0009$), 
suggesting that the WMAP K-band flux
densities are systematically lower than the ATCA flux densities by
a factor of $\simeq 1.2$. Using our measurements at the closer
frequency of 22 GHz (which has, however, a larger calibration
uncertainty) we find a median value $\log(S_{22{\rm GHz}}/S_K) =
0.057$ (mean $\log(S_{22{\rm GHz}}/S_K) = 0.101\pm 0.025$),
confirming the indication of a small, but non-negligible offset.

\begin{figure}
   \centering
\includegraphics[height=8cm, width=8.5cm]{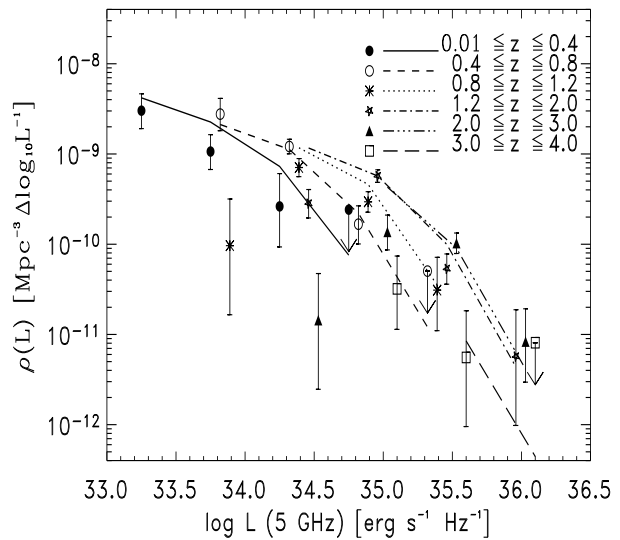}
\caption{Comoving luminosity function of FSRQs in the K\"uhr et
al. (1981) sample for different redshift bins (points), compared with the
model by De Zotti et al. 2005 (lines).  } \label{fig:kuhrlf}
\end{figure}

\section{The epoch-dependent luminosity function of FSRQs} \label{sec:lf}

\subsection{The 5 GHz luminosity function} \label{sec:5GHzlf}

To estimate the epoch-dependent 5 GHz luminosity function of FSRQs
we exploit the full K\"uhr et al. (1981) sample, including the
Northern portion, not observed at 18.5 GHz. The redshift
information is essentially complete: Stickel et al. (1994) list
redshifts for 198 of the 214 FSRQs in that sample and a search in 
the NASA-IPAC Extragalactic Database (NED) yielded redshift
measurements for 8 additional objects. In summary, the 
redshift completeness
is 94\% in the flux density bin $0\le \log S_5(\hbox{Jy}) < 0.2$,
98\% in the bin $0.2\le \log S_5(\hbox{Jy}) < 0.4$ and 100\% at
higher flux densities.

We have estimated the 5~GHz luminosity function, averaged over
intervals $\Delta \log L =0.5$, in six redshift bins ($0.01< z \le
0.4$,  $0.4 < z \leq 0.8$, $0.8 <z \leq 1.2$, $1.2 < z \leq 2$, $2
< z \leq 3$, $3 < z \leq 4$) using the classical $1/V_{\rm max}$
method (Schmidt 1968):
\begin{equation} \label{eqn:phi5}
\Phi(L_5,z)=\sum_{L_i\in [L_5-\Delta L/2,L_5+\Delta L/2]}\
\frac{w_i}{V_{\rm max,i}}
\end{equation}
where $V_{\rm max}$ is the maximum volume accessible to each
source and $w_i$ is the weight factor
correcting for the redshift incompleteness. $V_{\rm max}$ is given by the 
integral of the volume element from the lower
bound of the redshift bin to $z_{\rm max}= \hbox{min}(z_{\rm
up},z_{\rm lim})$, where $z_{\rm up}$ is the upper bound of the bin
and $z_{\rm lim}$ is the redshift at which the source would have a
flux density of 1 Jy (i.e. the lower limit of the K\"uhr sample). 
The weight factor $w_i$ is set at
$w_i=1/0.94$ and at $w_i=1/0.98$, respectively, for sources in the
two lowest flux density bins, defined above, and at $w_i=1$ for
brighter sources. The K-correction was computed using the spectral
index $\alpha^5_{2.7}$ of each source (see K\"uhr et al. 1981). 
The area covered by the
catalog is 9.81 sr. The 68\% confidence intervals on the
luminosity function were computed assuming Poisson statistics
with an effective number of sources per bin given by:
\begin{equation} \label{eqn:phi5err}
n_{\rm eff}=\left(\sum_i 1/V_{\rm max,i}\right)^2/\sum_i
1/V^2_{\rm max,i} \ .
\end{equation}
In Fig.~\ref{fig:kuhrlf} the estimated luminosity functions (points) are
compared with the model by De Zotti et al. (2005), averaged over
the same luminosity and redshift bins (lines). The agreement is generally
satisfactory. In spite of its simplicity (it assumes pure
luminosity evolution), the model correctly reproduces the positive
evolution up to $z\sim 2.5$ and the subsequent negative evolution.

\begin{figure}
   \centering
\includegraphics[height=6cm, width=8.5cm]{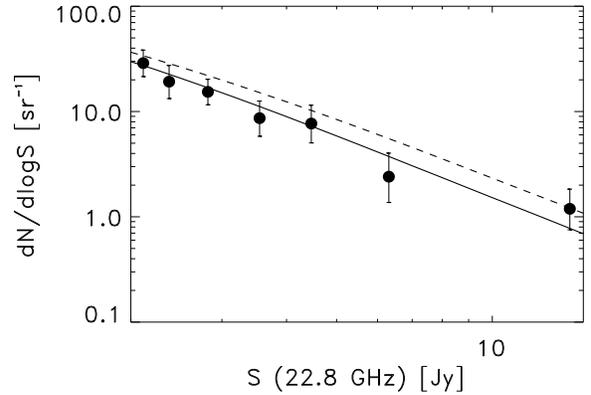}
 \caption{Differential counts of FSRQs in the
complete WMAP K-band sample (see text), compared with the
extrapolations of 5 GHz counts using the Gaussian representations
of the 5--18.5 GHz spectral index distribution of
Fig.~\protect\ref{fig:alphisto}, namely $<\alpha>=0.13$ and
$\sigma= 0.44$ (dashed line) or $\sigma= 0.30$ (solid line). The WMAP
fluxes have been multiplied by 1.2 (see
Sect.~\protect\ref{sec:analysis}). The error bars show the 68\%
confidence intervals for Poisson statistics (Gehrels 1986). }
\label{fig:wmapcounts}
\end{figure}

\begin{figure}
   \centering
\includegraphics[height=9cm, width=8.5cm]{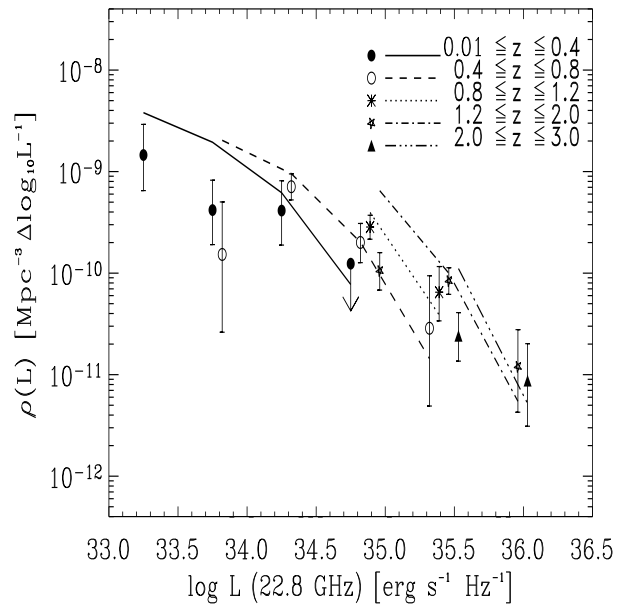}
\caption{Comoving luminosity function of FSRQs in the complete
WMAP sample (see text) for different redshift bins compared with
the model by De Zotti et al. (2005).} \label{fig:wmaplf}
\end{figure}

\subsection{Extrapolation to high frequencies} \label{sec:WMAPlf}

We now use the WMAP K-band catalogue to derive the 22.8~GHz luminosity
function and compare it to the one extrapolated from 5~GHz.
After having
multiplied the WMAP fluxes by 1.2 (see Sect.~\ref{sec:analysis}),
and using the 4.85~GHz flux densities from the GB6 (Gregory et al.
1996) or PMN (available at http://www.parkes.atnf.csiro.au/)
catalogs, 143 of them have been classified as flat-spectrum based
on the 4.85--22.8 GHz spectral indices ($\alpha_{4.85}^{22.8} <
0.5$, $S_\nu \propto \nu^{-\alpha}$). A search in the NASA/IPAC
NED database yielded optical identifications for 122 of the
flat-spectrum sources (94 FSRQs and 28 BL Lacs). The 21 objects
classified as galaxies or unidentified have been partitioned
randomly between FSRQs and BL Lacs, in proportion to the number of
identified objects in each class; this procedure has added 16
objects to the FSRQ sample.

In Fig.~\ref{fig:wmapcounts} the K-band differential counts of WMAP 
FSRQs (points) are compared with the ones extrapolated from the corresponding 
5~GHz counts, as described by the De Zotti et al. (2005) model, 
exploiting the Gaussian representations of the spectral index
distribution of Fig.~\ref{fig:alphisto}. If we adopt the formal
best-fit value of the dispersion, $\sigma=0.44$, we over-predict the WMAP 
counts (see dashed line in Fig.~\ref{fig:wmapcounts}) , confirming that the 
spectral index distribution is broadened by measurement errors. 
With $\sigma=0.30$, close to the dispersion of the 2.7--5 GHz spectral 
index distribution, the agreement is good (see solid line in 
Fig.~\ref{fig:wmapcounts}), indicating that high frequency 
surveys do not detect many FSRQs with ``anomalous'' radio spectra.

Redshift measurements were found in the literature for 99 of the
110 objects in the WMAP FSRQ sample. We have estimated their luminosity
function for the same luminosity and redshift bins as for the
K\"uhr sample (see Fig.~\ref{fig:kuhrlf}), except for the highest redshift 
bin, because we have only one WMAP source at $z>3$. The K-corrections have 
been computed using the median 4.85--18.5 GHz spectral index 
$\alpha_{4.85}^{18.5}=0.16$; this has
been preferred to $\alpha_{4.85}^{22.8}$ of individual objects
which are too liable to variability. 
In Fig.~\ref{fig:wmaplf}
these estimates (points) are compared with the extrapolations of the 5 GHz
luminosity function as modeled by the De Zotti et al. (2005),
which provides a good representation of the data (see lines), when we use
the 5--18.5 GHz Gaussian spectral
index distribution with $<\alpha>=0.13$ and $\sigma=0.30$.

\section{Conclusions} \label{sec:concls}

The ATCA 18.5 and 22~GHz flux density measurements of Southern
extragalactic sources in the complete 5~GHz sample of K\"uhr et
al. (1981) have been analyzed. The non-imaging technique used
yielded reliable flux density measurements for 66\% of
steep-spectrum sources. Although this incompleteness requires
caution in dealing with the data, some interesting indications
emerge. First the high frequency (5--18.5 GHz) spectral indices
are systematically steeper than the low frequency (2.7--5 GHz)
ones, the median steepening being $\Delta\alpha = 0.32$. There is
a hint of a larger steepening for sources classified as galaxies
than for those classified as QSOs. Since QSOs have generally
higher luminosities and higher redshifts than galaxies, this
difference translates into an anti-correlation of
$\Delta\alpha$ with the radio luminosity  (or $z$) when we consider the 
full steep-spectrum sample.
As a consequence, the {\it positive} correlation of the radio luminosity 
with the low frequency spectral index (see, e.g.,
Dunlop \& Peacock 1990), seems to {\it get reversed} when we consider the high
frequency spectral index. More complete data would however be 
necessary to test the reliability of this indication.

The information on 18.5 GHz flux densities of flat-spectrum
sources (mostly FSRQs) has a much better completeness level
(89\%). A high-frequency steepening is again present, with median
values $\alpha^5_{2.7}=-0.14$, $\alpha^{18.5}_{5}=0.16$,
$\Delta\alpha = 0.19$. Luminosities of these sources are
positively correlated, though weakly, with both the low and the
high frequency spectral indices.

In addition to having almost complete 18.5 GHz data, FSRQs have
almost complete redshift information, that has been exploited to
estimate their luminosity function in several redshift bins. The 5
GHz luminosity functions, computed using the full 1 Jy sample,
while confirming that the radio luminosity density peaks at
$z_{\rm peak} \simeq 2.5$ (Dunlop \& Peacock 1990; Shaver et al.
1996, 1999) do not provide evidence for deviations from pure
luminosity evolution, reported by other analyses (Hook et al.
1998; Vigotti et al. 2003), perhaps due to the poor statistics.
In particular we notice that the derived epoch-dependent 5~GHz luminosity 
function is fairly well represented by the model by De Zotti et al. (2005).
 
We exploit the multi-frequency 2.7, 5 and 18.5~GHz information, for Southern 
K\"uhr FSRQs to better extrapolate the De Zotti et al. model to 22.8~GHz, i.e. 
the central frequency of the WMAP K-band survey. This allows us to properly 
take into account the spectral steepening of FSQRs between 5 and 18.5~GHz.

The extrapolated model is compared to the sample of FSRQs detected in the
K-band by WMAP. To this respect, it is interesting to note that a direct 
comparison between ATCA 18.5/22~GHz and WMAP K-band flux density measurements 
for flat-spectrum sources hints towards a systematic under-estimation of WMAP 
fluxes by a factor $\sim 1.2$.

After correcting the WMAP K-band fluxes for the afore-mentioned offset, we are
able to get very good agreement between the differential counts of WMAP K-band 
FSRQs and the modeled extrapolation to 22.8~GHz. 
The same happens when we analyze the luminosity properties of
the FSRQ WMAP sample. The K-band epoch-dependent luminosity 
function derived from the sample of WMAP FSRQs is fairly well reproduced 
by the modeled 5~GHz luminosity function, extrapolated to 22.8 GHz
as described above.

Such results imply that the WMAP survey does not detect substantial numbers 
of FSRQs with anomalous spectra. 

\begin{acknowledgements}

This research was supported in part by the Italian Space Agency
(ASI) and by the Italian MIUR through a COFIN grant. RR warmly
thanks the Paul Wild Observatory staff for their kind hospitality
at Narrabri (NSW, Australia) where some of this work was
accomplished. The Australia Telescope is funded by the
Commonwealth of Australia for operation as a National Facility
managed by CSIRO. This research has made use of the NASA/IPAC
Extragalactic Database (NED) which is operated by the Jet
Propulsion Laboratory, California Institute of Technology, under
contract with the National Aeronautics and Space Administration.

\end{acknowledgements}

% references

\setcounter{table}{0}
\begin{table*}
\begin{center}
\caption{18.5 and 22~GHz flux density measurements of Southern sources in the
K\"uhr sample. The flag in the last column indicates the reliability of the 
18.5~GHz measurements. Also listed are the 5~GHz flux density and the 
$2.7 - 5$~GHz spectral index (from K\"uhr et al. 1981), the source type and 
redshift (from Stickel et al. 1994, plus some redshifts from NED).} 
\label{tab:zS5}
\begin{tabular}{rrrccrrrrc} \hline\hline
\multicolumn{1}{c}{Name} & $S_{5\rm{GHz}}$ & $\alpha_{2.7}^5$ &
type & $z$ &
\multicolumn{1}{c}{$S_{18.5\rm{GHz}}$} &  \multicolumn{1}{c}{$S_{22\rm{GHz}}$} & ext \\
       & \multicolumn{1}{c}{(Jy)} & & & & \multicolumn{1}{c}{(Jy)} & \multicolumn{1}{c}{(Jy)} & \\\hline
 0003-066 &   1.48 &  -0.02 &    QSO &   0.3470 & $  2.352 \pm  0.118 $ & $  2.551 \pm 0.255  $ & - \\
 0003-003 &   1.39 &   0.84 &    QSO &   1.0370 & $  0.474 \pm  0.028 $ & $  0.454 \pm 0.045  $ & - \\
 0008-421 &   1.23 &   1.14 &    GAL &   0.1716 & $  0.182 \pm  0.009 $ & $  0.134 \pm 0.013  $ & - \\
 0022-423 &   1.83 &   0.72 &    QSO &   0.9370 & $  0.339 \pm  0.017 $ & $  0.252 \pm 0.025  $ & - \\
 0023-263 &   3.62 &   0.72 &    GAL &   0.3220 & $  1.129 \pm  0.057 $ & $  1.051 \pm 0.105  $ & - \\
 0034-014 &   1.59 &   0.73 &    GAL &   0.0730 & $  0.211 \pm  0.011 $ & $  0.176 \pm 0.018  $ & R \\
 0035-024 &   2.69 &   0.65 &    GAL &   0.2197 & $  0.652 \pm  0.035 $ & $  0.608 \pm 0.061  $ & - \\
 0039-445 &   1.21 &   0.88 &    GAL &   0.3460 & $  0.197 \pm  0.010 $ & $  0.175 \pm 0.018  $ & - \\
 0042-357 &   1.01 &   0.79 &  --- &      --  & $  0.218 \pm  0.011 $ & $    0.215 \pm 0.022  $ & - \\
 0043-424 &   3.17 &   0.72 &    GAL &   0.0526 & $  0.165 \pm  0.008 $ & $  0.162 \pm 0.016  $ & - \\
 0045-255 &   2.34 &   0.74 &    GAL &   0.0009 & $  0.571 \pm  0.029 $ & $  0.489 \pm 0.049  $ & - \\
 0047-579 &   2.40 &  -0.33 &    QSO &   1.7970 & $  1.647 \pm  0.082 $ & $  1.037 \pm 0.104  $ & - \\
 0048-097 &   1.98 &  -0.43 & BL/QSO &   0.1221 & $  1.384 \pm  0.069 $ & $  1.144 \pm 0.114  $ & - \\
 0055-016 &   2.13 &   0.79 &    GAL &   0.0447 & $  0.108 \pm  0.006 $ & $  0.091 \pm 0.009  $ & R \\
 0056-572 &   1.10 &  -0.58 &    QSO &   0.0180 & $  0.580 \pm  0.029 $ & $  0.777 \pm 0.078  $ & - \\
 0105-163 &   1.18 &   1.00 &    GAL &   0.4000 & $  0.084 \pm  0.004 $ & $  0.060 \pm 0.006  $ & - \\
 0112-017 &   1.20 &   0.01 &    QSO &   1.3810 & $  0.789 \pm  0.040 $ & $  1.186 \pm 0.119  $ & - \\
 0113-118 &   1.94 &  -0.14 &    QSO &   0.6720 & $  1.419 \pm  0.071 $ & $  1.440 \pm 0.144  $ & - \\
 0114-476 &   2.31 &   0.60 &    GAL &   0.1460 & $  0.017 \pm  0.002 $ & $  0.030 \pm 0.003  $ & R \\
 0114-211 &   1.28 &   0.89 &    GAL &   1.4100 & $  0.283 \pm  0.014 $ & $  0.243 \pm 0.024  $ & - \\
 0117-155 &   1.61 &   0.87 &    GAL &   0.5650 & $  0.267 \pm  0.013 $ & $  0.154 \pm 0.015  $ & - \\
 0118-272 &   1.14 &  -0.28 & BL/QSO &   0.5570 & $  0.827 \pm  0.042 $ & $  0.850 \pm 0.085  $ & R \\
 0122-003 &   1.23 &   0.02 &    QSO &   1.0700 & $  1.728 \pm  0.087 $ & $  1.664 \pm 0.166  $ & - \\
 0123-016 &   1.85 &   0.83 &    GAL &   0.0177 & $  0.156 \pm  0.008 $ & $  0.154 \pm 0.015  $ & R \\
 0130-171 &   1.00 &   0.06 &    QSO &   1.0220 & $  1.541 \pm  0.078 $ & $  1.675 \pm 0.168  $ & - \\
 0131-522 &   1.22 &  -0.18 &    QSO &   0.0200 & $  0.797 \pm  0.040 $ & $  1.045 \pm 0.104  $ & - \\
 0131-367 &   2.23 &   1.49 &    GAL &   0.0300 & $  0.054 \pm  0.003 $ & $  0.068 \pm 0.007  $ & - \\
 0135-247 &   1.70 &  -0.35 &    QSO &   0.8310 & $  0.873 \pm  0.044 $ & $  1.705 \pm 0.171  $ & R \\
 0138-097 &   1.22 &  -0.88 & BL/QSO &   0.5010 & $  0.929 \pm  0.047 $ & $  1.110 \pm 0.111  $ & - \\
 0159-117 &   1.39 &   0.59 &    QSO &   0.6690 & $  0.681 \pm  0.035 $ & $  0.772 \pm 0.077  $ & - \\
 0201-440 &   1.04 &   0.80 &   QSO? &      --  & $  0.257 \pm  0.013 $ & $  0.238 \pm 0.024  $ & - \\
 0202-172 &   1.38 &   0.03 &    QSO &   1.7400 & $  1.076 \pm  0.054 $ & $  1.523 \pm 0.152  $ & - \\
 0208-512 &   3.31 &   0.12 &    QSO &   1.0030 & $  2.893 \pm  0.145 $ & $  6.142 \pm 0.614  $ & - \\
 0213-132 &   1.71 &   0.81 &    GAL &   0.1480 & $  0.183 \pm  0.009 $ & $  0.133 \pm 0.013  $ & R \\
 0235-197 &   1.45 &   0.83 &    GAL &   0.6200 & $  0.179 \pm  0.009 $ & $  0.160 \pm 0.016  $ & - \\
 0237-233 &   3.40 &   0.68 &    QSO &   2.2240 & $  0.919 \pm  0.046 $ & $  0.848 \pm 0.085  $ & - \\
 0238-084 &   1.44 &  -1.19 &    GAL &   0.0050 & $  1.014 \pm  0.052 $ & $  1.622 \pm 0.162  $ & - \\
 0240-002 &   1.80 &   0.86 &    GAL &   0.0034 & $  0.432 \pm  0.022 $ & $  0.342 \pm 0.034  $ & R \\
 0252-712 &   1.63 &   1.04 &    GAL &   0.5660 & $  0.317 \pm  0.016 $ & $  0.204 \pm 0.020  $ & - \\
 0302-623 &   1.41 &  -0.18 &    QSO &      --  & $  1.866 \pm  0.093 $ & $  1.468 \pm 0.147  $ & - \\
 0308-611 &   1.36 &  -0.76 &    QSO &      --  & $  1.211 \pm  0.061 $ & $  1.141 \pm 0.114  $ & - \\
 0319-453 &   1.02 &   1.14 &    GAL &   0.0633 & $  0.009 \pm  0.002 $ & $  0.017 \pm 0.002  $ & R \\ 
 0332-403 &   2.56 &  -0.43 &    QSO &   1.4450 & $  1.781 \pm  0.089 $ & $  1.631 \pm 0.163  $ & - \\
 0336-019 &   2.86 &  -0.30 &    QSO &   0.8520 & $  4.062 \pm  0.203 $ & $  3.262 \pm 0.326  $ & - \\
 0344-345 &   1.39 &   0.68 &    GAL &   0.0538 & $  0.098 \pm  0.005 $ & $  0.040 \pm 0.004  $ & R \\
 0349-278 &   2.34 &   0.51 &    GAL &   0.0656 & $  0.025 \pm  0.002 $ & $  0.040 \pm 0.004  $ & R \\
 0400-319 &   1.06 &   0.12 &    QSO &   1.2880 & $  0.876 \pm  0.044 $ & $  0.711 \pm 0.071  $ & - \\
 0402-362 &   1.38 &  -0.46 &    QSO &   1.4170 & $  3.175 \pm  0.159 $ & $  1.311 \pm 0.131  $ & - \\
 0403-132 &   2.88 &   0.05 &    QSO &   0.5710 & $  2.281 \pm  0.114 $ & $  2.731 \pm 0.273  $ & - \\
 0405-385 &   1.09 &  -0.11 &    QSO &   1.2850 & $  1.638 \pm  0.082 $ & $  1.154 \pm 0.115  $ & - \\
 0405-123 &   1.96 &   0.30 &    QSO &   0.5740 & $  1.413 \pm  0.071 $ & $  1.185 \pm 0.118  $ & - \\
 0407-658 &   3.43 &   1.13 &   QSO? &      --  & $  0.461 \pm  0.023 $ & $  0.349 \pm 0.035  $ & - \\
 0409-752 &   4.43 &   0.87 &    GAL &   0.6940 & $  0.906 \pm  0.045 $ & $  0.726 \pm 0.073  $ & - \\
 0413-210 &   1.43 &   0.29 &    QSO &   0.8070 & $  0.858 \pm  0.043 $ & $  1.195 \pm 0.120  $ & - \\
 0414-189 &   1.35 &  -0.22 &    QSO &   1.5360 & $  0.904 \pm  0.045 $ & $  0.549 \pm 0.055  $ & - \\
\hline
\end{tabular}
\end{center}
\end{table*}

\setcounter{table}{0}
\begin{table*}
\begin{center}
\caption{Continued. }
\begin{tabular}{rrrccrrrrc} \hline\hline
\multicolumn{1}{c}{Name} & $S_{5\rm{GHz}}$ & $\alpha_{2.7}^5$ &
type & $z$ &
\multicolumn{1}{c}{$S_{18.5\rm{GHz}}$} &  \multicolumn{1}{c}{$S_{22\rm{GHz}}$} & ext \\
       & \multicolumn{1}{c}{(Jy)} & & & & \multicolumn{1}{c}{(Jy)} & \multicolumn{1}{c}{(Jy)} & \\\hline
 0420-014 &   1.46 &  -0.01 &    QSO &   0.9150 & $  7.931 \pm  0.397 $ & $  7.111 \pm 0.711  $ & - \\
 0426-380 &   1.17 &  -0.20 & BL/QSO &   1.0300 & $  1.053 \pm  0.053 $ & $  1.206 \pm 0.121  $ & - \\
 0427-539 &   2.32 &   0.54 &    GAL &   0.0390 & $  0.055 \pm  0.003 $ & $  0.043 \pm 0.004  $ & - \\
 0434-188 &   1.23 &  -0.25 &    QSO &   2.7020 & $  0.424 \pm  0.021 $ & $  0.436 \pm 0.044  $ & - \\
 0437-454 &   1.41 &  -0.24 &    QSO &      --  & $  1.124 \pm  0.056 $ & $  1.088 \pm 0.109  $ & - \\
 0438-436 &   6.94 &  -0.25 &    QSO &   2.8520 & $  3.896 \pm  0.195 $ & $  3.512 \pm 0.351  $ & - \\
 0440-003 &   2.61 &   0.29 &    QSO &   0.8440 & $  1.215 \pm  0.061 $ & $  1.408 \pm 0.141  $ & - \\
 0442-282 &   2.20 &   0.89 &    GAL &   0.1470 & $  0.086 \pm  0.006 $ & $  0.286 \pm 0.029  $ & R \\
 0454-810 &   1.40 &  -0.29 &    QSO &   0.4440 & $  1.565 \pm  0.078 $ & $  2.033 \pm 0.203  $ & - \\
 0451-282 &   2.26 &   0.04 &    QSO &   2.5590 & $  1.747 \pm  0.087 $ & $  2.565 \pm 0.257  $ & - \\
 0453-206 &   1.84 &   0.70 &    GAL &   0.0354 & $  0.158 \pm  0.008 $ & $  0.106 \pm 0.011  $ & R \\
 0454-463 &   2.32 &   0.03 &    QSO &   0.8580 & $  3.585 \pm  0.179 $ & $  2.532 \pm 0.253  $ & - \\
 0454-234 &   2.06 &  -0.16 &    QSO &   1.0030 & $  5.401 \pm  0.270 $ & $  2.495 \pm 0.249  $ & - \\
 0458-020 &   1.74 &   0.08 &    QSO &   2.2860 & $  1.570 \pm  0.079 $ & $  1.224 \pm 0.122  $ & - \\
 0506-612 &   1.73 &   0.10 &    QSO &   1.0930 & $  2.455 \pm  0.123 $ & $  2.742 \pm 0.274  $ & - \\
 0511-484 &   1.81 &   0.63 &    GAL &   0.3063 & $  0.068 \pm  0.004 $ & $  0.100 \pm 0.010  $ & R \\
 0511-220 &   1.31 &  -0.13 &    QSO &   1.2960 & $  0.768 \pm  0.039 $ & $  0.603 \pm 0.060  $ & - \\
 0514-459 &   1.06 &   0.32 &    QSO &   0.1940 & $  0.887 \pm  0.044 $ & $  1.156 \pm 0.116  $ & - \\
 0518-458 &  15.45 &   1.02 &    GAL &   0.0342 & $  0.780 \pm  0.039 $ & $  1.096 \pm 0.110  $ & - \\
 0521-365 &   9.29 &   0.43 &    GAL &   0.0550 & $  2.777 \pm  0.139 $ & $  3.971 \pm 0.397  $ & R \\
 0524-460 &   1.02 &  -0.14 &    QSO &   1.4790 & $  0.524 \pm  0.026 $ & $  0.779 \pm 0.078  $ & - \\
 0528-250 &   1.16 &   0.20 &    QSO &   2.7650 & $  0.468 \pm  0.024 $ & $  0.390 \pm 0.039  $ & - \\
 0537-441 &   4.00 &  -0.06 & BL/QSO &   0.8960 & $ 10.667 \pm  0.533 $ & $  9.031 \pm 0.903  $ & - \\
 0537-286 &   1.02 &  -0.52 &    QSO &   3.1190 & $  0.865 \pm  0.043 $ & $  1.639 \pm 0.164  $ & - \\
 0539-057 &   1.55 &  -1.41 &    QSO &   0.8390 & $  0.757 \pm  0.038 $ & $  0.913 \pm 0.091  $ & - \\
 0602-319 &   1.25 &   0.67 &    QSO &   0.4520 & $  0.417 \pm  0.021 $ & $  0.262 \pm 0.026  $ & - \\
 0604-203 &   1.04 &   0.89 &    GAL &   0.1640 & $  0.105 \pm  0.006 $ & $  0.061 \pm 0.006  $ & - \\
 0605-085 &   3.49 &  -0.09 &    QSO &   0.8720 & $  1.930 \pm  0.097 $ & $  2.456 \pm 0.246  $ & - \\
 0606-223 &   1.40 &  -0.55 &    QSO &   1.9260 & $  0.951 \pm  0.048 $ & $  0.896 \pm 0.090  $ & - \\
 0607-157 &   1.82 &   0.01 &    QSO &   0.3240 & $  5.991 \pm  0.300 $ & $  5.451 \pm 0.545  $ & - \\
 0614-349 &   1.37 &   0.58 &    GAL &   0.3290 & $  0.497 \pm  0.025 $ & $  0.433 \pm 0.043  $ & - \\
 0618-371 &   1.37 &   0.40 &    GAL &   0.0326 & $  0.024 \pm  0.005 $ & $  0.035 \pm 0.004  $ & R \\
 0620-526 &   1.27 &   0.79 &    GAL &   0.0511 & $  0.243 \pm  0.012 $ & $  0.176 \pm 0.018  $ & R \\
 0625-536 &   1.86 &   1.12 &    GAL &   0.0540 & $  0.077 \pm  0.004 $ & $  0.080 \pm 0.008  $ & R \\
 0625-354 &   2.20 &   0.45 &    GAL &   0.0550 & $  1.097 \pm  0.055 $ & $  0.792 \pm 0.079  $ & R \\
 0637-752 &   5.85 &  -0.19 &    QSO &   0.6540 & $  4.780 \pm  0.239 $ & $  4.118 \pm 0.412  $ & - \\
 0642-349 &   1.02 &  -0.13 &    QSO &   2.1650 & $  0.322 \pm  0.016 $ & $  0.352 \pm 0.035  $ & - \\
 0743-673 &   1.79 &   0.69 &    QSO &   1.5110 & $  1.556 \pm  0.078 $ & $  1.384 \pm 0.138  $ & R \\
 0743-006 &   1.99 &  -0.57 &    QSO &   0.9940 & $  1.614 \pm  0.081 $ & $  1.317 \pm 0.132  $ & - \\
 0805-077 &   1.04 &   0.32 &    QSO &   1.8370 & $  1.661 \pm  0.083 $ & $  2.288 \pm 0.229  $ & - \\
 0806-103 &   1.64 &   0.67 &    GAL &   0.1070 & $  0.189 \pm  0.010 $ & $  0.308 \pm 0.031  $ & - \\
 0825-202 &   1.18 &   0.93 &    QSO &   0.8220 & $  0.217 \pm  0.011 $ & $  0.361 \pm 0.036  $ & R \\
 0834-201 &   3.72 &  -0.03 &    QSO &   2.7520 & $  4.353 \pm  0.218 $ & $  1.982 \pm 0.198  $ & - \\
 0834-196 &   1.52 &   0.92 &   GAL? &   1.0320 & $  0.410 \pm  0.021 $ & $  0.444 \pm 0.044  $ & - \\
 0842-754 &   1.42 &   0.67 &    QSO &   0.5240 & $  0.308 \pm  0.016 $ & $  0.509 \pm 0.051  $ & - \\
 0858-279 &   1.42 &   0.55 &    QSO &   2.1520 & $  1.459 \pm  0.073 $ & $  1.247 \pm 0.125  $ & - \\
 0859-257 &   1.74 &   0.92 &    GAL &   0.3050 & $  0.182 \pm  0.009 $ & $  0.343 \pm 0.034  $ & R \\
 0859-140 &   2.30 &   0.42 &    QSO &   1.3390 & $  1.190 \pm  0.060 $ & $  1.334 \pm 0.133  $ & - \\
 0915-118 &  13.99 &   0.86 &    GAL &   0.0547 & $  2.169 \pm  0.109 $ & $  1.803 \pm 0.180  $ & - \\
 0919-260 &   1.32 &  -0.22 &    QSO &   2.3000 & $  1.441 \pm  0.072 $ & $  1.441 \pm 0.144  $ & - \\
 0941-080 &   1.09 &   0.72 &    GAL &   0.2280 & $  0.369 \pm  0.019 $ & $  0.425 \pm 0.043  $ & - \\
 1015-314 &   1.40 &   0.75 &    QSO &   1.3460 & $  0.454 \pm  0.023 $ & $  0.437 \pm 0.044  $ & - \\
 1017-426 &   1.27 &   0.98 &    QSO &   1.2800 & $  0.251 \pm  0.013 $ & $  0.177 \pm 0.018  $ & - \\
 1032-199 &   1.15 &  -0.07 &    QSO &   2.1980 & $  1.261 \pm  0.063 $ & $  1.282 \pm 0.128  $ & - \\
 1045-188 &   1.14 &  -0.32 &    QSO &   0.5950 & $  1.626 \pm  0.081 $ & $  2.966 \pm 0.297  $ & - \\
 1046-409 &   1.07 &   0.41 &    QSO &   0.6200 & $  0.413 \pm  0.021 $ & $  0.418 \pm 0.042  $ & R \\
 1057-797 &   1.62 &  -0.58 &    QSO &      --  & $  2.561 \pm  0.128 $ & $  2.286 \pm 0.229  $ & - \\
\hline
\end{tabular}
\end{center}
\end{table*}

\setcounter{table}{0}
\begin{table*}
\begin{center}
\caption{Continued. }
\begin{tabular}{rrrccrrrrc} \hline\hline
\multicolumn{1}{c}{Name} & $S_{5\rm{GHz}}$ & $\alpha_{2.7}^5$ &
type & $z$ &
\multicolumn{1}{c}{$S_{18.5\rm{GHz}}$} &  \multicolumn{1}{c}{$S_{22\rm{GHz}}$} & ext \\
       & \multicolumn{1}{c}{(Jy)} & & & & \multicolumn{1}{c}{(Jy)} & \multicolumn{1}{c}{(Jy)} & \\\hline
 1104-445 &   2.07 &  -0.16 &    QSO &   1.5980 & $  2.278 \pm  0.114 $ & $  3.040 \pm 0.304  $ & - \\
 1116-462 &   1.35 &   0.35 &    QSO &   0.7130 & $  0.959 \pm  0.048 $ & $  0.911 \pm 0.091  $ & - \\
 1127-145 &   6.57 &  -0.03 &    QSO &   1.1870 & $  2.926 \pm  0.146 $ & $  3.099 \pm 0.310  $ & - \\
 1136-135 &   2.11 &   0.42 &    QSO &   0.5540 & $  0.434 \pm  0.022 $ & $  0.430 \pm 0.043  $ & R \\
 1143-483 &   1.23 &   0.64 &    GAL &   0.2982 & $  0.418 \pm  0.021 $ & $  0.612 \pm 0.061  $ & - \\
 1143-245 &   1.18 &   0.18 &    QSO &   1.9500 & $  0.692 \pm  0.035 $ & $  0.620 \pm 0.062  $ & - \\
 1144-379 &   1.61 &  -0.66 & BL/QSO &   1.0480 & $  4.433 \pm  0.222 $ & $  3.193 \pm 0.319  $ & - \\
 1145-071 &   1.25 &  -0.22 &    QSO &   1.3420 & $  0.811 \pm  0.041 $ & $  0.541 \pm 0.054  $ & - \\
 1148-001 &   1.90 &   0.44 &    QSO &   1.9820 & $  0.964 \pm  0.048 $ & $  0.880 \pm 0.088  $ & - \\
 1151-348 &   2.83 &   0.65 &    QSO &   0.2580 & $  0.992 \pm  0.050 $ & $  0.857 \pm 0.086  $ & - \\
 1202-262 &   1.04 &   0.41 &    QSO &   0.7890 & $  0.514 \pm  0.026 $ & $  0.506 \pm 0.051  $ & - \\
 1213-172 &   1.47 &  -0.05 &    GAL &   0.6691 & $  1.897 \pm  0.095 $ & $  2.068 \pm 0.207  $ & - \\
 1221-423 &   1.03 &   0.76 &    GAL &   0.1706 & $  0.357 \pm  0.018 $ & $  0.347 \pm 0.035  $ & - \\
 1229-021 &   1.07 &   0.29 &    QSO &   1.0450 & $  0.676 \pm  0.034 $ & $  0.492 \pm 0.049  $ & - \\
 1237-101 &   1.31 &   0.25 &    QSO &   0.7530 & $  1.035 \pm  0.052 $ & $  1.014 \pm 0.101  $ & - \\
 1239-044 &   1.01 &   1.05 &    GAL &   0.4800 & $  0.184 \pm  0.009 $ & $  0.290 \pm 0.029  $ & - \\
 1243-072 &   1.03 &  -0.44 &    QSO &   1.2860 & $  0.700 \pm  0.035 $ & $  0.608 \pm 0.061  $ & - \\
 1244-255 &   1.36 &  -0.30 &    QSO &   0.6330 & $  1.848 \pm  0.093 $ & $  1.416 \pm 0.142  $ & - \\
 1245-197 &   2.50 &   0.70 &    QSO &   1.2750 & $  0.764 \pm  0.038 $ & $  0.593 \pm 0.059  $ & - \\
 1246-410 &   1.38 &   0.77 &    GAL &   0.0093 & $  0.213 \pm  0.011 $ & $  0.206 \pm 0.021  $ & - \\
 1251-122 &   2.76 &   0.83 &    GAL &   0.0145 & $  0.032 \pm  0.002 $ & $  0.081 \pm 0.008  $ & R \\
 1253-055 &  14.95 &  -0.30 &    QSO &   0.5360 & $ 27.946 \pm  1.397 $ & $ 20.602 \pm 2.060  $ & - \\
 1255-316 &   1.73 &  -0.24 &    QSO &   1.9240 & $  1.996 \pm  0.100 $ & $  1.377 \pm 0.138  $ & - \\
 1302-102 &   1.17 &  -0.17 &    QSO &   0.2860 & $  0.820 \pm  0.041 $ & $  0.969 \pm 0.097  $ & - \\
 1306-095 &   1.94 &   0.68 &   GAL? &   0.4640 & $  0.827 \pm  0.042 $ & $  0.696 \pm 0.070  $ & - \\
 1308-220 &   1.16 &   1.23 &    GAL &   0.8000 & $  0.216 \pm  0.011 $ & $  0.284 \pm 0.028  $ & - \\
 1313-333 &   1.36 &  -0.50 &    QSO &   1.2100 & $  1.457 \pm  0.073 $ & $  4.031 \pm 0.403  $ & - \\
 1318-434 &   2.02 &   0.69 &    GAL &   0.0110 & $  0.534 \pm  0.027 $ & $  0.569 \pm 0.057  $ & - \\
 1320-446 &   1.07 &   0.83 &    QSO &      --  & $  0.283 \pm  0.014 $ & $  0.236 \pm 0.024  $ & - \\
 1322-428 &   3.00 &   0.35 &    GAL &   0.0016 & $  6.479 \pm  0.324 $ & $  6.631 \pm 0.663  $ & - \\
 1331-098 &   1.33 &   0.60 &    GAL &   0.0810 & $  0.068 \pm  0.006 $ & $  0.472 \pm 0.047  $ & R \\
 1333-337 &   6.90 &   0.60 &    GAL &   0.0129 & $  0.267 \pm  0.013 $ & $  0.349 \pm 0.035  $ & - \\
 1334-127 &   2.24 &  -0.17 &    QSO &   0.5390 & $  7.036 \pm  0.352 $ & $ 11.007 \pm 1.101  $ & - \\
 1335-061 &   1.02 &   0.95 &    QSO &   0.6250 & $  0.150 \pm  0.008 $ & $  0.271 \pm 0.027  $ & - \\
 1352-104 &   1.01 &  -0.40 &    QSO &   0.3320 & $  1.623 \pm  0.081 $ & $  0.913 \pm 0.091  $ & - \\
 1354-152 &   1.52 &   0.10 &    QSO &   1.8900 & $  0.998 \pm  0.050 $ & $  1.126 \pm 0.113  $ & - \\
 1355-416 &   1.44 &   0.88 &    QSO &   0.3130 & $  0.101 \pm  0.005 $ & $  0.135 \pm 0.014  $ & - \\
 1406-076 &   1.08 &  -0.19 &    QSO &   1.4940 & $  1.771 \pm  0.089 $ & $  1.170 \pm 0.117  $ & - \\
 1424-418 &   3.13 &  -0.28 &    QSO &   1.5220 & $  2.913 \pm  0.146 $ & $  1.593 \pm 0.159  $ & - \\
 1451-375 &   1.89 &  -0.37 &    QSO &   0.3140 & $  2.022 \pm  0.101 $ & $  2.026 \pm 0.203  $ & - \\
 1453-109 &   1.52 &   0.77 &    QSO &   0.9380 & $  0.427 \pm  0.022 $ & $  0.342 \pm 0.034  $ & R \\
 1504-166 &   1.98 &   0.16 &    QSO &   0.8760 & $  1.447 \pm  0.073 $ & $  2.334 \pm 0.233  $ & - \\
 1508-055 &   2.43 &   0.30 &    QSO &   1.1910 & $  1.282 \pm  0.064 $ & $  0.837 \pm 0.084  $ & - \\
 1510-089 &   3.08 &  -0.31 &    QSO &   0.3610 & $  2.490 \pm  0.125 $ & $  2.089 \pm 0.209  $ & - \\
 1514-241 &   2.00 &   0.16 & BL/GAL &   0.0486 & $  1.911 \pm  0.096 $ & $  2.351 \pm 0.235  $ & - \\
 1519-273 &   2.35 &  -0.27 & BL/QSO &   0.0710 & $  1.586 \pm  0.080 $ & $  1.140 \pm 0.114  $ & - \\
 1524-136 &   1.23 &   0.63 &    QSO &   1.6870 & $  0.459 \pm  0.023 $ & $  0.443 \pm 0.044  $ & - \\
 1541-828 &   1.47 &  -0.20 &   QSO? &      --  & $  0.589 \pm  0.030 $ & $  0.500 \pm 0.050  $ & - \\
 1550-269 &   1.17 &   0.23 &    QSO &   2.1450 & $  0.584 \pm  0.030 $ & $  0.538 \pm 0.054  $ & R \\
 1547-795 &   1.40 &   0.79 &    GAL &   0.4830 & $  0.150 \pm  0.008 $ & $  0.125 \pm 0.013  $ & R \\
 1549-790 &   3.67 &   0.26 &    GAL &   0.1490 & $  1.434 \pm  0.072 $ & $  1.160 \pm 0.116  $ & - \\
 1555+001 &   2.24 &  -0.34 &    QSO &   1.7700 & $  0.880 \pm  0.044 $ & $  0.688 \pm 0.069  $ & - \\
 1602-093 &   1.19 &   0.92 &    GAL &   0.1090 & $  0.078 \pm  0.006 $ & $  0.073 \pm 0.007  $ & R \\
 1610-771 &   3.91 &  -0.06 &    QSO &   1.7100 & $  2.654 \pm  0.133 $ & $  2.252 \pm 0.225  $ & - \\
 1619-680 &   1.86 &  -0.07 &    QSO &   1.3540 & $  0.718 \pm  0.036 $ & $  0.581 \pm 0.058  $ & - \\
 1622-253 &   2.08 &   0.14 &    QSO &   0.7860 & $  2.200 \pm  0.112 $ & $  0.988 \pm 0.099  $ & - \\
 1637-771 &   2.87 &   0.56 &    GAL &   0.0427 & $  0.424 \pm  0.021 $ & $  0.503 \pm 0.050  $ & - \\ 
\hline
\end{tabular}
\end{center}
\end{table*}

\setcounter{table}{0}
\begin{table*}
\begin{center}
\caption{Continued. }
\begin{tabular}{rrrccrrrrc} \hline\hline
\multicolumn{1}{c}{Name} & $S_{5\rm{GHz}}$ & $\alpha_{2.7}^5$ &
type & $z$ &
\multicolumn{1}{c}{$S_{18.5\rm{GHz}}$} &  \multicolumn{1}{c}{$S_{22\rm{GHz}}$} & ext \\
       & \multicolumn{1}{c}{(Jy)} & & & & \multicolumn{1}{c}{(Jy)} & \multicolumn{1}{c}{(Jy)} & \\\hline
 1655-776 &   1.13 &   0.38 &    GAL &   0.0944 & $  0.123 \pm  0.008 $ & $  0.505 \pm 0.051  $ & R \\
 1717-009 &  22.15 &   0.78 &    GAL &   0.0304 & $  0.150 \pm  0.012 $ & $  0.190 \pm 0.019  $ & R \\
 1718-649 &   3.81 &   0.08 &    GAL &   0.0145 & $  3.002 \pm  0.150 $ & $  2.498 \pm 0.250  $ & - \\
 1721-026 &   1.19 &   0.41 &    GAL &   0.0330 & $  0.020 \pm  0.001 $ & $  0.063 \pm 0.006  $ & R \\
 1733-565 &   3.45 &   0.46 &    GAL &   0.0985 & $  0.216 \pm  0.011 $ & $  0.198 \pm 0.020  $ & - \\
 1737-608 &   1.14 &   0.62 &   GAL? &   0.3689 & $  0.072 \pm  0.004 $ & $  0.061 \pm 0.006  $ & - \\ 
 1741-038 &   3.68 &  -0.75 &    QSO &   1.0570 & $  4.873 \pm  0.245 $ & $  3.723 \pm 0.372  $ & - \\
 1740-517 &   3.04 &   0.67 &   GAL? &   0.4016 & $  1.369 \pm  0.069 $ & $  1.096 \pm 0.110  $ & - \\
 1814-637 &   4.48 &   0.75 &    GAL &   0.0627 & $  1.581 \pm  0.079 $ & $  1.383 \pm 0.138  $ & - \\
 1815-553 &   1.35 &   0.02 &    QSO &      --  & $  1.088 \pm  0.054 $ & $  0.779 \pm 0.078  $ & - \\
 1829-718 &   1.02 &   1.03 &  --- &      --  & $  0.258 \pm  0.013 $ & $    0.197 \pm 0.020  $ & - \\
 1831-711 &   1.19 &   0.17 &    QSO &   1.3560 & $  2.222 \pm  0.111 $ & $  1.688 \pm 0.169  $ & - \\
 1839-486 &   1.30 &   0.68 &    GAL &   0.1120 & $  0.196 \pm  0.010 $ & $  0.093 \pm 0.009  $ & - \\
 1903-802 &   1.84 &  -0.29 &    QSO &   0.5000 & $  0.615 \pm  0.031 $ & $  0.534 \pm 0.053  $ & - \\
 1929-397 &   1.02 &   0.69 &    GAL &   0.0746 & $  0.038 \pm  0.003 $ & $  0.029 \pm 0.003  $ & R \\
 1932-464 &   3.46 &   1.05 &    GAL &   0.2310 & $  0.507 \pm  0.025 $ & $  0.411 \pm 0.041  $ & R \\
 1933-400 &   1.48 &  -0.26 &    QSO &   0.9660 & $  1.165 \pm  0.058 $ & $  1.245 \pm 0.124  $ & - \\
 1934-638 &   6.26 &   0.96 &    GAL &   0.1830 & $  1.169 \pm  0.059 $ & $  0.813 \pm 0.081  $ & - \\
 1936-155 &   1.69 &  -0.26 &    QSO &      --  & $  0.968 \pm  0.049 $ & $  1.118 \pm 0.112  $ & - \\
 1938-155 &   2.24 &   0.96 &    GAL &   0.4520 & $  0.494 \pm  0.025 $ & $  0.417 \pm 0.042  $ & - \\
 1936-623 &   1.10 &   0.16 &    QSO &   0.6982 & $  0.535 \pm  0.027 $ & $  0.512 \pm 0.051  $ & - \\
 1954-388 &   2.06 &  -0.43 &    QSO &   0.6260 & $  3.800 \pm  0.190 $ & $  2.734 \pm 0.273  $ & - \\
 1954-552 &   2.38 &   0.76 &    GAL &   0.0598 & $  0.298 \pm  0.015 $ & $  0.209 \pm 0.021  $ & - \\
 1958-179 &   1.20 &  -0.13 &    QSO &   0.6500 & $  1.364 \pm  0.069 $ & $  0.800 \pm 0.080  $ & - \\
 2000-330 &   1.15 &  -0.78 &    QSO &   3.7770 & $  0.516 \pm  0.026 $ & $  0.517 \pm 0.052  $ & - \\
 2005-489 &   1.23 &  -0.24 & BL/GAL &   0.0710 & $  1.339 \pm  0.067 $ & $  0.927 \pm 0.093  $ & - \\
 2008-068 &   1.37 &   0.68 &    GAL &   0.5470 & $  0.401 \pm  0.021 $ & $  0.308 \pm 0.031  $ & R \\
 2008-159 &   1.39 &  -1.02 &    QSO &   1.1800 & $  2.269 \pm  0.114 $ & $  1.100 \pm 0.110  $ & - \\
 2020-575 &   1.08 &   0.71 &    GAL &   0.3520 & $  0.101 \pm  0.005 $ & $  0.072 \pm 0.007  $ & R \\
 2032-350 &   1.94 &   0.89 &   GAL? &   1.0682 & $  0.395 \pm  0.020 $ & $  0.337 \pm 0.034  $ & - \\
 2037-253 &   1.20 &  -0.42 &    QSO &   1.5740 & $  0.447 \pm  0.024 $ & $  0.890 \pm 0.089  $ & R \\
 2044-027 &   1.02 &   0.52 &    QSO &   0.9420 & $  0.360 \pm  0.019 $ & $  0.296 \pm 0.030  $ & - \\
 2053-201 &   1.02 &   0.69 &    GAL &   0.1560 & $  0.097 \pm  0.007 $ & $  0.193 \pm 0.019  $ & R \\
 2052-474 &   2.52 &   0.11 &    QSO &   1.4910 & $  0.741 \pm  0.037 $ & $  1.441 \pm 0.144  $ & - \\
 2058-282 &   2.13 &   0.62 &    GAL &   0.0377 & $  0.096 \pm  0.005 $ & $  0.124 \pm 0.012  $ & R \\
 2104-256 &   4.55 &   0.73 &    GAL &   0.0370 & $  0.078 \pm  0.004 $ & $  0.087 \pm 0.009  $ & R \\
 2106-413 &   2.35 &  -0.17 &    QSO &   1.0547 & $  1.891 \pm  0.095 $ & $  2.290 \pm 0.229  $ & - \\
 2126-158 &   1.28 &  -0.14 &    QSO &   3.2660 & $  0.931 \pm  0.048 $ & $  0.790 \pm 0.079  $ & - \\
 2128-123 &   2.07 &  -0.11 &    QSO &   0.5010 & $  3.611 \pm  0.181 $ & $  3.000 \pm 0.300  $ & - \\
 2131-021 &   2.12 &  -0.13 & BL/QSO &   0.5570 & $  2.244 \pm  0.114 $ & $  1.693 \pm 0.169  $ & - \\
 2135-147 &   1.41 &   0.70 &    QSO &   0.2000 & $  0.152 \pm  0.010 $ & $  0.500 \pm 0.050  $ & R \\ 
 2135-209 &   1.55 &   0.77 &    GAL &   0.6350 & $  0.451 \pm  0.024 $ & $  0.580 \pm 0.058  $ & - \\
 2142-758 &   1.32 &   0.07 &    QSO &   1.1390 & $  0.685 \pm  0.034 $ & $  0.625 \pm 0.062  $ & - \\
 2140-816 &   1.00 &   0.73 &  --- &      --  & $  0.114 \pm  0.006 $ & $    0.086 \pm 0.009  $ & - \\
 2149-287 &   1.36 &   0.61 &    GAL &   0.4790 & $  0.411 \pm  0.021 $ & $  0.410 \pm 0.041  $ & - \\
 2150-520 &   1.21 &   0.95 &   QSO? &      --  & $  0.259 \pm  0.013 $ & $  0.193 \pm 0.019  $ & - \\
 2152-699 &  12.28 &   0.63 &    GAL &   0.0285 & $  1.412 \pm  0.071 $ & $  1.248 \pm 0.125  $ & R \\
 2155-152 &   1.77 &  -0.15 &    QSO &   0.6720 & $  2.472 \pm  0.125 $ & $  1.480 \pm 0.148  $ & - \\
 2203-188 &   4.38 &   0.28 &    QSO &   0.6190 & $  2.197 \pm  0.111 $ & $  2.000 \pm 0.200  $ & - \\
 2204-540 &   2.41 &  -0.16 &    QSO &   1.2060 & $  1.277 \pm  0.064 $ & $  1.175 \pm 0.117  $ & - \\
 2206-237 &   1.01 &   0.45 &    GAL &   0.0870 & $  0.340 \pm  0.017 $ & $  0.284 \pm 0.028  $ & - \\
 2210-257 &   1.05 &  -0.15 &    QSO &   1.8330 & $  0.604 \pm  0.034 $ & $  1.100 \pm 0.110  $ & - \\
 2211-172 &   2.17 &   1.21 &    GAL &   0.1530 & $  0.059 \pm  0.003 $ & $  0.178 \pm 0.018  $ & R \\
 2216-038 &   1.50 &  -0.48 &    QSO &   0.9010 & $  2.540 \pm  0.128 $ & $  1.818 \pm 0.182  $ & - \\
 2221-023 &   2.11 &   0.76 &    GAL &   0.0562 & $  0.039 \pm  0.002 $ & $  0.045 \pm 0.005  $ & R \\
 2223-052 &   4.51 &   0.11 &    QSO &   1.4040 & $  8.129 \pm  0.407 $ & $  7.253 \pm 0.725  $ & - \\
 2226-411 &   1.08 &   0.87 &    QSO &   0.4462 & $  0.293 \pm  0.015 $ & $  0.254 \pm 0.025  $ & - \\ 
\hline
\end{tabular}
\end{center}
\end{table*}

\setcounter{table}{0}
\begin{table*}
\begin{center}
\caption{Continued. }
\begin{tabular}{rrrccrrrrc} \hline\hline
\multicolumn{1}{c}{Name} & $S_{5\rm{GHz}}$ & $\alpha_{2.7}^5$ &
type & $z$ &
\multicolumn{1}{c}{$S_{18.5\rm{GHz}}$} &  \multicolumn{1}{c}{$S_{22\rm{GHz}}$} & ext \\
       & \multicolumn{1}{c}{(Jy)} & & & & \multicolumn{1}{c}{(Jy)} & \multicolumn{1}{c}{(Jy)} & \\\hline
 2227-088 &   1.77 &   0.03 &    QSO &   1.5610 & $  1.618 \pm  0.081 $ & $  1.499 \pm 0.150  $ & - \\
 2227-399 &   1.05 &  -0.05 &    QSO &   0.3230 & $  0.754 \pm  0.038 $ & $  0.582 \pm 0.058  $ & R \\
 2240-260 &   1.03 &   0.08 & BL/QSO &   0.7740 & $  0.573 \pm  0.029 $ & $  0.430 \pm 0.043  $ & R \\
 2243-123 &   2.45 &   0.18 &    QSO &   0.6300 & $  2.183 \pm  0.110 $ & $  2.200 \pm 0.220  $ & - \\
 2245-328 &   1.85 &   0.13 &    QSO &   2.2550 & $  0.164 \pm  0.008 $ & $  0.227 \pm 0.023  $ & R \\
 2250-412 &   1.33 &   0.92 &    GAL &   0.3100 & $  0.262 \pm  0.013 $ & $  0.183 \pm 0.018  $ & - \\
 2252-530 &   1.01 &   0.94 &   GAL? &      --  & $  0.274 \pm  0.014 $ & $  0.238 \pm 0.024  $ & - \\
 2255-282 &   1.78 &  -0.41 &    QSO &   0.9260 & $ 11.255 \pm  0.563 $ & $  2.250 \pm 0.225  $ & - \\
 2311-452 &   1.47 &   0.38 &    QSO &   2.8840 & $  0.578 \pm  0.029 $ & $  0.602 \pm 0.060  $ & - \\
 2317-277 &   1.30 &   0.64 &    GAL &   0.1730 & $  0.076 \pm  0.004 $ & $  0.045 \pm 0.005  $ & - \\
 2323-407 &   1.09 &   0.79 &    GAL &   0.2982 & $  0.214 \pm  0.011 $ & $  0.167 \pm 0.017  $ & - \\
 2324-023 &   1.16 &   0.49 &    GAL &   0.1880 & $  0.292 \pm  0.016 $ & $  0.278 \pm 0.028  $ & R \\
 2326-477 &   2.33 &   0.06 &    QSO &   1.3060 & $  1.531 \pm  0.077 $ & $  1.094 \pm 0.109  $ & - \\
 2329-162 &   1.06 &   0.06 &    QSO &   1.1550 & $  1.996 \pm  0.100 $ & $  1.044 \pm 0.104  $ & - \\
 2331-240 &   1.09 &  -0.08 &    GAL &   0.0477 & $  1.151 \pm  0.058 $ & $  0.679 \pm 0.068  $ & - \\
 2331-417 &   1.59 &   0.88 &    GAL &   0.9070 & $  0.256 \pm  0.013 $ & $  0.253 \pm 0.025  $ & - \\
 2333-528 &   1.22 &   0.09 &    QSO &      --  & $  1.059 \pm  0.053 $ & $  0.946 \pm 0.095  $ & - \\
 2337-334 &   1.21 &   0.20 &    QSO &   1.8020 & $  0.502 \pm  0.025 $ & $  0.593 \pm 0.059  $ & R \\
 2345-167 &   3.66 &  -0.51 &    QSO &   0.5760 & $  2.412 \pm  0.121 $ & $  1.407 \pm 0.141  $ & - \\
 2353-686 &   1.10 &  -0.26 &    QSO &   1.7160 & $  0.785 \pm  0.039 $ & $  0.502 \pm 0.050  $ & - \\
 2354-117 &   1.48 &   0.24 &    QSO &   0.9600 & $  1.197 \pm  0.060 $ & $  0.925 \pm 0.093  $ & - \\
 2355-534 &   1.52 &  -0.36 &    QSO &   1.0060 & $  1.523 \pm  0.076 $ & $  1.605 \pm 0.161  $ & - \\
 2356-611 &   7.73 &   0.80 &    GAL &   0.0958 & $  0.060 \pm  0.003 $ & $  0.212 \pm 0.021  $ & R \\ \hline
\end{tabular}
\end{center}
\end{table*}


\normalsize
\begin{thebibliography}{99}

%\bibitem{} Avni, Y., \& Bahcall, J.N., 1980, ApJ, 235, 694

\bibitem[\protect\citeauthoryear{Arg{\" u}eso, Gonz{\' a}lez-Nuevo, \&
Toffolatti}{2003}]{2003ApJ...598...86A} Arg{\" u}eso, F., Gonz{\'
a}lez-Nuevo, J., \& Toffolatti, L. 2003, ApJ, 598, 86

\bibitem[\protect\citeauthoryear{Bennett et
al.}{2003}]{2003ApJS..148...97B} Bennett, C.L., Hill, R.S.,
Hinshaw, G., et al. 2003, ApJS, 148, 97

%\bibitem{} Carroll, S.M., Press, W.H., \& Turner, E.L., 1992,
%Annu. Rev. Astron. Astrophys., 30, 499

\bibitem{} De Zotti, G., Ricci, R., Mesa, D., Silva, L., Mazzotta, P.,
Toffolatti, L., \& Gonz\'alez-Nuevo, J. 2005, A\&A, 431, 893

\bibitem{} Dunlop, J.S. \& Peacock, J.A. 1990, MNRAS, 247, 19

%\bibitem{} Efstathiou, G., Ellis, R.S., \& Peterson, B.A., 1988, MNRAS, 232,
%431

\bibitem{} Feigelson, E.D., \& Nelson, P.I., 1985, ApJ, 293, 192

%\bibitem{} Gregory, P.C., Vavasour, J.D., Scott, W.K., Condon, J.J.,
%1994, ApJS, 90, 173

\bibitem{} Gehrels, N. 1986, ApJ, 303, 336

\bibitem{} Gregory, P.C., Scott, W.K.,
Douglas, K., \& Condon, J.J. 1996, ApJS, 103, 427

\bibitem{} Hook, I.M., Shaver, P.A., \& McMahon, R.G. 1998,
in {\it The Young Universe}, ASP Conf. series, S. D'Odorico, A.
Fontana, E. Giallongo, eds., Vol. 146, p. 17

\bibitem{} Isobe, T., \& Feigelson, E.D., 1990, Bull. Amer. Astro. Society, 
22, 917
\bibitem{} Isobe, T., Feigelson, E.D., \& Nelson, P.I., 1986, ApJ, 306, 490

\bibitem{} K\"uhr, H., Witzel, A., Pauliny-Toth, I.I.K., \& Nauber, U. 1981,
A\&AS, 45, 367

%\bibitem{} Loveday, J., 2000, MNRAS, 312, 557

\bibitem{} Ricci, R., Prandoni, I., Gruppioni, C., Sault, R.J., \&
De Zotti, G. 2004a, A\&A, 415, 549 (Paper~I)

\bibitem{} Ricci, R., Sadler, E.M., Ekers, R.D., Staveley--Smith, L., Wilson,
W.E., Kesteven, M.J., Subrahmanyan, R., Walker, M.A., Jackson,
C.A., \& De Zotti, G. 2004b, MNRAS, 354, 305

\bibitem{} Schmidt, M. 1968, ApJ, 151, 393

%\bibitem{} Schmidt, M., Schneider, D.P., \& Gunn, J.E. 1995, AJ, 110, 68

\bibitem{} Shaver, P.A., Wall, J.V., Kellermann, K.I., Jackson,
C.A., \& Hawkins, M.R.S. 1996, Natur, 384, 439

\bibitem[\protect\citeauthoryear{Shaver et
al.}{1999}]{1999ASPC..156..163S} Shaver, P.A., Hook, I.M.,
Jackson, C.A., Wall, J.V., Kellermann, K.I. 1999, in Highly
Redshifted Radio Lines, ASP Conf. Series, C.L. Carilli, S.J.E.
Radford, K.M. Menten, \& G.I. Langston eds., 156, 163

%\bibitem{} Sondr\'e, L., \& Lahav, O., 1993, MNRAS, 260, 285

\bibitem{} Stickel, M., Meisenheimer, K., \& K\"uhr, H. 1994, A\&A Supp.
Ser., 105, 211

\bibitem[\protect\citeauthoryear{Vigotti et
al.}{2003}]{2003ApJ...591...43V} Vigotti, M., Carballo, R., Benn,
C.R., De Zotti, G., Fanti, R., Gonzalez Serrano, J.I., Mack,
K.-H., \& Holt, J. 2003, ApJ, 591, 43


\bibitem{} Waldram, E.M., Pooley, G.G., Grainge, K.J.B., Jones, M.E., 
Saunders, R.D.E., Scott, P.F., Taylor, A.C., 2003, MNRAS, 342, 915

%\bibitem{} Warren, S.J., Hewett, P.C., Osmer, P.S., 1994, ApJ, 421, 412

\end{thebibliography}
\end{document}